\documentclass[preprint,12pt]{elsarticle}
\newcommand{\bc}{\color{black}}
\newcommand{\bb}{\color{black}}

\usepackage{graphicx,amssymb,amsmath,color,numcompress}
\journal{Physics Letters B}

\begin{document}

\begin{frontmatter}

\title{Superscaling predictions for neutrino-induced charged-current charged pion production at MiniBooNE}

\author[label1,label2]{M.~V.~Ivanov\corref{cor}}
\cortext[cor]{Corresponding author.}
\ead{martin.inrne@gmail.com}
\author[label1]{J.~M.~Udias}
\author[label2]{A.~N.~Antonov}
\author[label3]{J.~A. Caballero}
\author[label4]{M.~B.~Barbaro}
\author[label1]{E.~Moya de Guerra}

\address[label1]{Grupo de F\'{i}sica Nuclear, Departamento de F\'{i}sica At\'omica, Molecular y Nuclear, Facultad de Ciencias F\'{i}sicas, Universidad Complutense de Madrid, CEI Moncloa, Madrid E-28040, Spain}
\address[label2]{Institute for Nuclear Research and Nuclear Energy, Bulgarian Academy of Sciences, Sofia 1784, Bulgaria}
\address[label3]{Departamento de F\'{i}sica At\'omica, Molecular y Nuclear, Universidad de Sevilla, 41080 Sevilla, Spain}
\address[label4]{Dipartimento di Fisica, Universit\`{a} di Torino and INFN, Sezione di Torino, Via P. Giuria 1, 10125 Torino, Italy}

\begin{abstract}

Superscaling approximation (SuSA) predictions to neutrino-induced charged-current charged pion production in the $\Delta$-resonance region are explored under MiniBooNE experimental conditions. The results obtained within SuSA for the flux-averaged double-differential cross sections of the $\pi^+$ production for the  $\nu_\mu+\text{CH}_2$ reaction as a function of the muon kinetic energy and of the scattering angle, the cross sections averaged over the angle, the total cross section for the $\pi^+$ production, as well as CC1$\pi^+$ to CCQE cross section ratio are compared with the corresponding MiniBooNE experimental data. {The SuSA predictions are in good agreement with data on neutrino flux average cross-sections, but a somewhat different dependence on the neutrino energy is predicted than the one resulting from the experimental analysis.}

\end{abstract}


\end{frontmatter}

\section{Introduction \label{sec:1}}

The properties of neutrinos, particularly the parameters of their oscillations, are being studied with increasing interest as these may carry important information about the limits of the Standard Model. In most neutrino experiments, the interactions of the neutrinos occur with nucleons bound in nuclei. The influence of nucleon-nucleon interactions on the response of nuclei to neutrino probes must then be considered, ideally in a model independent way. Model predictions for these reactions involve many different effects such as nuclear correlations, interactions in the final state, possible modification of the nucleon properties inside the nuclear medium, that presently cannot be computed in an unambiguous and precise way. This is particularly true for the channels where neutrino interactions take place by means of excitation of a nucleon resonance and ulterior production of mesons. The data on neutrino-induced charged-current (CC) charged and neutral pion production cross sections on mineral oil recently released by the MiniBooNE collaboration~\cite{neutr_pi+} provides an unprecedented opportunity to carry out a systematic study of double differential cross section of the processes, $\nu_\mu \, p \to \mu^- p \,\pi^+$, $\nu_\mu \, n \to \mu^- n \, \pi^+$, $\nu_\mu \, n \to \mu^-  p \, \pi^0$, averaged over the neutrino flux.

One way of avoiding model-dependencies is to use the nuclear response to other leptonic probes, such as  electrons, under similar conditions to the neutrino experiments. Thus, in this paper we compare the SuSA predictions for neutrino-induced CC charged pion production cross sections with MiniBooNE data~\cite{neutr_pi+}. The extensive analyses of scaling~\cite{neutr_sick80,neutr_Ciofi,neutr_Day90} and superscaling~\cite{neutr_Alberico90, neutr_Barbaro98, neutr_Donnelly99, neutr_Maieron02, neutr_Barbaro04, neutr_Antonov} phenomena observed in electron-nucleus scattering lead to the use of the scaling function directly extracted from $(e,e')$ data to predict neutrino (antineutrino)-nucleus cross sections~\cite{neutr_Amaro04}, not relying on a particular nuclear structure model. Within SuSA a {\sl ``superscaling function''} $f(\psi)$ is built by factoring-out the single-nucleon content off the double-differential cross section and plotting the remaining nuclear response versus a scaling variable $\psi(q,\omega)$. Approximate scaling of the first kind, {\it i.e.,} no explicit dependence of $f(\psi)$ on the momentum transfer $q$, can be seen at transfer energies below the quasielastic (QE) peak. Scaling of second kind, {\it i.e.,} no dependence of $f(\psi)$ on the mass number, turns out to be excellent in the same region. When scaling of both first and second types occur, one says that superscaling takes place.

The analyses of the world data on inclusive electron-nucleus scattering~\cite{neutr_Donnelly99} confirmed the observation of superscaling and thus
justified the extraction of a universal nuclear response to be also used for weak interacting probes. However, while there is a number of theoretical models that exhibit superscaling, such as for instance the relativistic Fermi gas (RFG)~\cite{neutr_Alberico90,neutr_Barbaro98}, the nuclear response departs from the one derived from the experimental data. This showed the necessity to consider more complex dynamical pictures of finite nuclear systems -- beyond the RFG -- in order to describe the nuclear response at intermediate energies. SuSA predictions are based on the phenomenological superscaling function extracted from the world data on quasielastic electron scattering~\cite{neutr_Jourdan96}.
{\bc The model has been  applied to neutral current scattering~\cite{neutr_Amaro06} and it has also been extended to the $\Delta$-resonance region~\cite{neutr_Amaro04} where the response of the nuclear system proceeds through excitation of {internal} nucleonic degrees of freedom. {Indeed, a non-quasielastic cross section for the excitation region in which nucleon excitations, particularly the $\Delta$, play a major role was obtained by subtracting from the data QE-equivalent cross sections given by SuSA~\cite{neutr_Barbaro04,NQE2}. This procedure has been possible due to the large amount of available high-quality data of inelastic electron scattering cross sections on $^{12}$C, including also separate information on the longitudinal and transverse responses, the latter containing important contributions introduced by effects beyond the impulse approximation (non-nucleonic).}

The SuSA procedure has been already employed to describe the non pionic (QE) cross-section of the MiniBooNE $\nu$- and $\overline\nu$-nucleus cross-section~\cite{neutr_Amaro10,neutr_Amaro11,Amaro:2011aa}. Here we extend the analysis to CC pion production cross-section measured at MiniBooNe, that from the theoretical point of view can be seen as more challenging. For instance, $\Delta$ properties in the nuclear medium, as well as both coherent and incoherent pion production for the nucleus should be considered in any theoretical approach, while in the SuSA procedure they are included phenomenologically extracted from the electron scattering data. All what is assumed within SuSA approach is an internal factorization of the nuclear response to a weakly interacting probe into a single-nucleon part and a `nuclear function' accounting for the {overall} interaction among nucleons. As mentioned before, the SuSA assumptions have been tested against a great deal of electron-nucleus scattering data with fair success. The factorization assumption allows to apply the same nuclear responses derived from electron scattering to neutrino-induced reactions, with a mere use of the adequate single-nucleon terms for this case. To show the importance of nuclear interaction effects as predicted within SuSA, as a reference, we also show results obtained within the RFG, with no interactions among nucleons, for which  the scaling function} in the $\Delta $-domain is simply given as  $f^{\Delta}_{RFG}(\psi_{\Delta})=\dfrac{3}{4}(1-{\psi_{\Delta}}^2)\theta(1-{\psi_{\Delta}}^2)$ with $\psi_{\Delta}$ the dimensionless scaling variable extracted from the RFG analysis that incorporates the typical momentum scale for the selected nucleus~\cite{neutr_Maieron02,neutr_Amaro04}. 

\begin{figure}[t]
  \centering
\includegraphics[width=0.85\columnwidth]{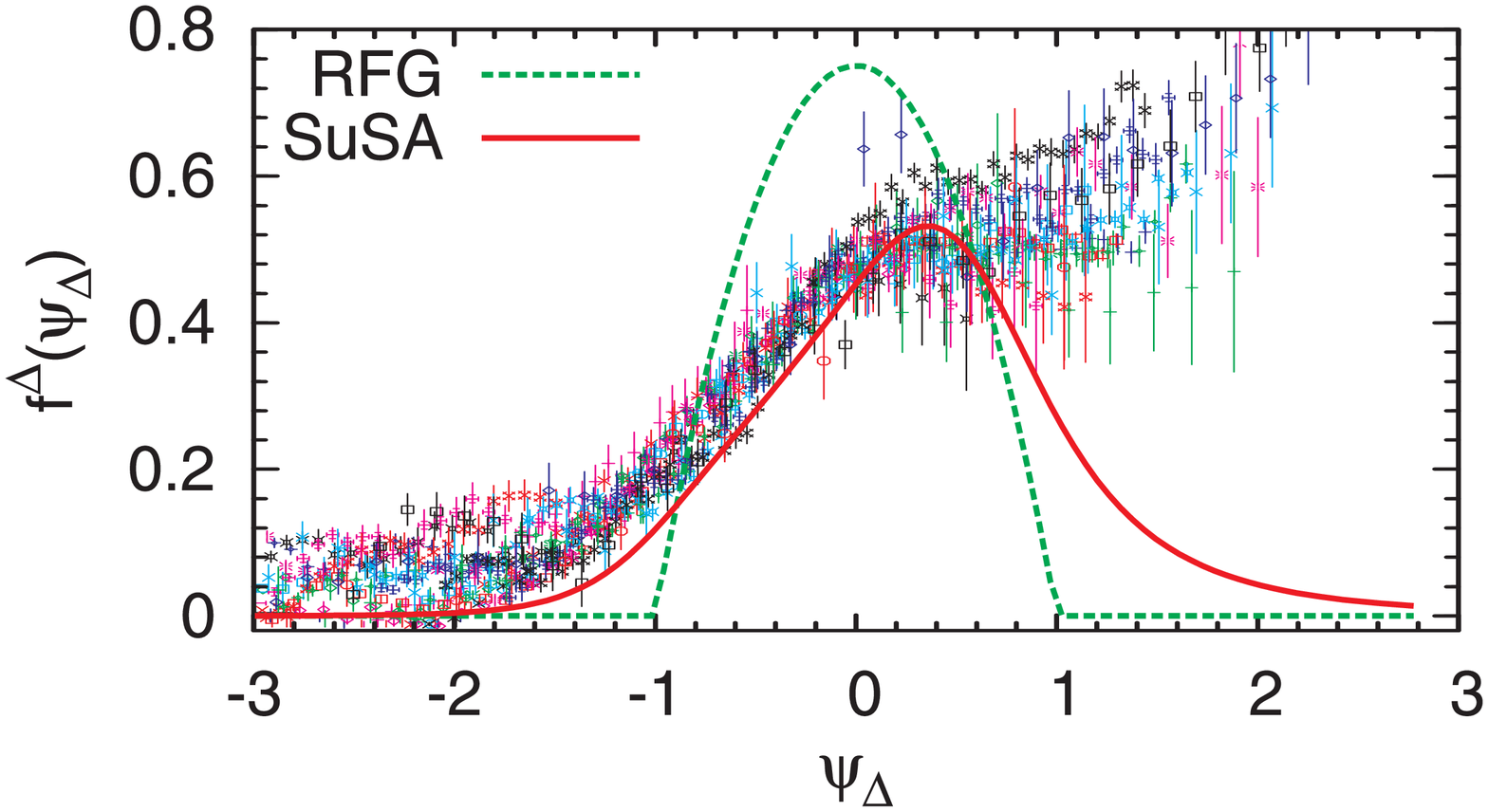}
\caption{(Color online) The SuSA scaling function in the $\Delta$-region $f^{\Delta}(\psi_{\Delta})$ (solid line)
extracted from the world data on electron scattering~\cite{neutr_Amaro04}. The dotted line shows the scaling
functions $f^\Delta(\psi_\Delta)$ in the RFG model.} \label{fig1}
\end{figure}

{\bb In Fig.~\ref{fig1} we compare the $\Delta$-region SuSA~\cite{neutr_Amaro04} and RFG scaling functions, that we use in our study. Here the data refer to $^{12}$C and $^{16}$O and span a large range of energies (from $0.3$ to $4$~GeV) and scattering angles (from $12$ to $145$ degrees). The experimental points in Fig.~\ref{fig1} are extracted by subtracting from the total cross sections the quasielastic contribution calculated using the universal QE scaling function $f_{\text{QE}}(\psi)$. This analysis does not include the contribution associated to meson-exchange currents (MEC), which are important in the region between the QE and $\Delta$ peaks and are responsible of the disagreement between the data and the fit (red curve) at large negative values of $\psi_\Delta$. These currents are mediated by virtual pions and do not correspond to the production of real pions; hence they should be included in the ``quasielastic'' response, which for the MiniBooNE experiment corresponds to the absence of real pions in the final state but not to the pionic data discussed in this work. The contribution of the MEC to the QE neutrino and antineutrino scattering has been evaluated in the SuSA framework in two recent papers~\cite{neutr_Amaro10,Amaro:2011aa}. We also note that the fit shown in Fig.~\ref{fig1} is restricted only to excitation energies at and below the $\Delta$ resonance peak, where the response is dominated by the $\Delta$; at higher energies other resonances and eventually the tail of deep inelastic scattering (DIS) contribute. This explains the difference between the phenomenological superscaling function, which aims to describe the $\Delta$ resonance peak, and the data observed in Fig.~\ref{fig1} at positive $\psi_{\Delta}$-values.}

\section{Formalism \label{sec:2}}

In what follows we present the results of applying the SuSA and RFG $\Delta$-scaling function to neutrino-induced CC charged pion production. We follow the formalism given in~\cite{neutr_Amaro04}. The charged current neutrino cross section in the target laboratory frame is given in the form
\begin{equation}
\frac{d^2 \sigma }{d\Omega dk' }=
\frac{(G \cos \theta_c k')^{2}}{2\pi ^{2}}
\left(1-\frac{|Q^2|}{4\epsilon\epsilon ^{\prime }}\right)
{\cal F}^{2}
\label{eq32ant}
\end{equation}
where $\Omega$, $k'$ and $\epsilon^\prime$ are the scattering angle, momentum and energy of the outgoing muon, $G$ is the Fermi constant and $\theta_c$ is the Cabibbo angle. The function ${\cal F}^{2}$ depends on the nuclear structure through the $R$ responses and can be written as~\cite{neutr_Amaro04,neutr_Umino95}:
\begin{equation*}
{\cal F}^{2}=\widehat{V}_\text{CC}R_\text{CC}+2\widehat{V}_\text{CL}R_\text{CL}
+\widehat{V}_\text{LL}R_\text{LL}+\widehat{V}_\text{T}R_\text{T}
+2\widehat{V}_\text{T$'$}R_\text{T$'$}\label{new33}
\end{equation*}
that is, as a generalized Rosenbluth decomposition having charge-charge (CC), charge-longitudinal (CL), longitudinal-longitudinal (LL) and two types of transverse (T,T$'$) responses ($R$'s) with the corresponding leptonic kinematical factors ($V$'s). The nuclear response functions in $\Delta$-region are expressed in terms of the nuclear tensor $W^{\mu\nu}$ in the corresponding region. The basic expressions used to calculate the single-nucleon cross sections are given in~\cite{neutr_Amaro04}. These involve the leptonic and hadronic tensors as well as the response and structure functions for single nucleons. A convenient parametrization of the single-nucleon $W^+n\to \Delta^+$ vertex is given in terms
of eight form-factors: four vector ($C^V_{3,4,5,6}$) and four axial ($C^A_{3,4,5,6}$) ones. Vector form factors have been determined from the analysis of photo and electro-production data, mostly on a deuteron target. Among the axial form factors, the most important contribution comes from $C^A_5$. The factor $C^A_6$, whose contribution to the differential cross section vanishes for massless leptons, can be related to $C^A_5$ by PCAC. Since there are no other theoretical constraints for $C^A_{3,4,5}(q^2)$, they have to be fitted to data. We use two different parameterizations: the one given in \cite{neutr_Alvarez-Ruso99} where deuteron effects were evaluated (authors estimated that the latter reduce the cross section by 10\%), denoted as ``PR1'', and the one from~\cite{neutr_Paschos04}, called ``PR2''.

With these ingredients, we evaluate the cross section for CC $\Delta^{++}$ and $\Delta^+$ production on proton and neutron, respectively. Once produced, the $\Delta$ decays into $\pi N$ pairs. For the amplitudes $\mathcal{A}$ of pion production the following isospin decomposition applies:
$\mathcal{A}(\nu_l \, p \to l^- p \,\pi^+)= \mathcal{A}_3$, 
$\mathcal{A}(\nu_l \, n \to l^- n \,\pi^+)= \frac{1}{3}\mathcal{A}_3 + \frac{2 \sqrt{2}}{3} \mathcal{A}_1$, 
$\mathcal{A}(\nu_l \, n \to l^-  p \, \pi^0)= -\frac{\sqrt{2}}{3}\mathcal{A}_3 + \frac{2}{3} \mathcal{A}_1$, 
with $\mathcal{A}_3$ being the amplitude for the isospin $3/2$ state of the $\pi N$ system, predominantly $\Delta$, and $\mathcal{A}_1$ the amplitude for the isospin $1/2$ state that is not considered here.

\begin{figure}[t]
\includegraphics[width=\columnwidth]{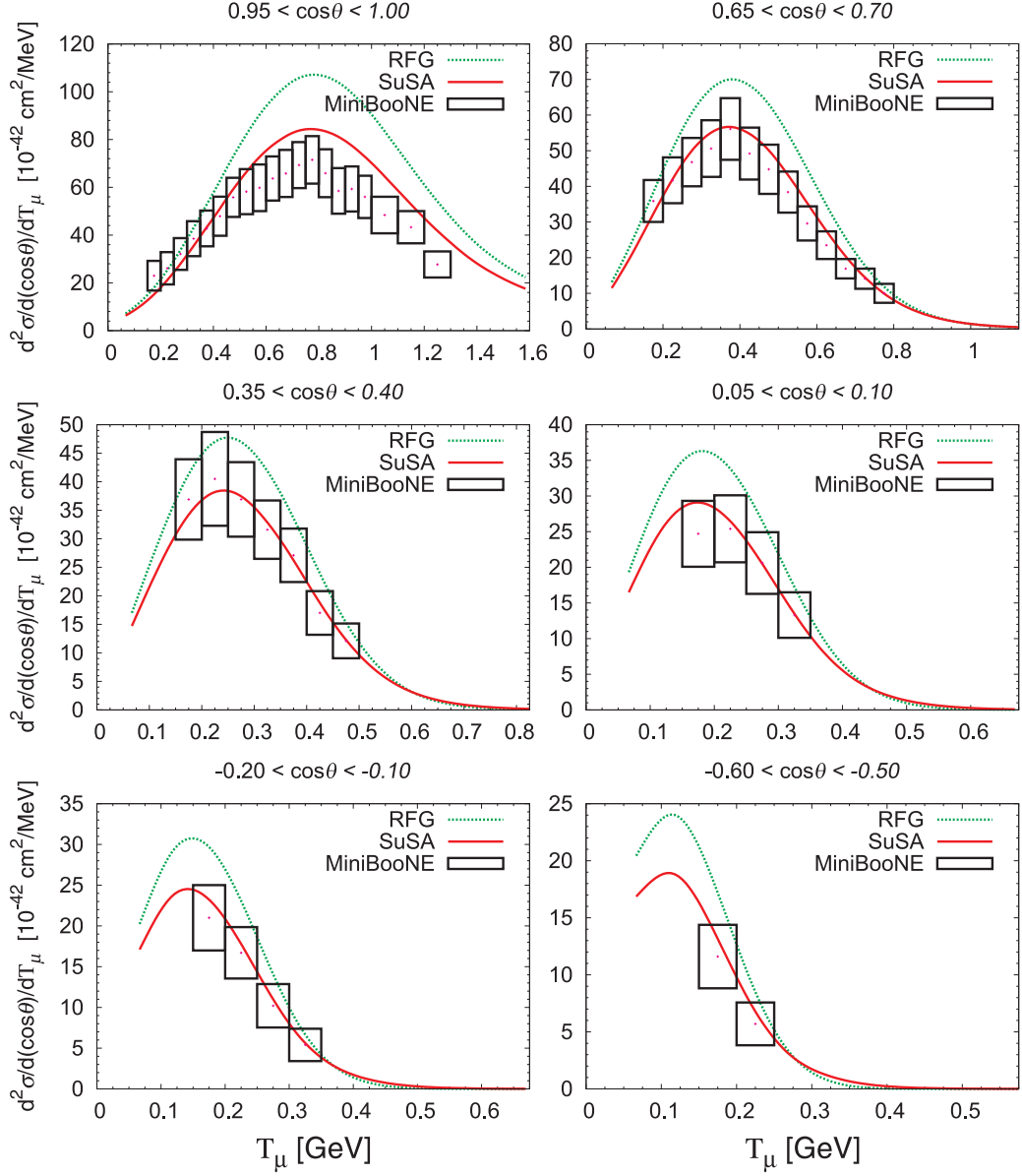}
\caption{(Color online) The double-differential cross section averaged over the neutrino energy flux as a function of
the muon kinetic energy $T_\mu$ obtained by SuSA (solid line) and RFG (dotted line) $\Delta$-region scaling functions.
In each subfigure the results have been averaged over the corresponding angular bin of $\cos\theta$. ``PR2''
parametrization~\cite{neutr_Paschos04} is used.
The results are compared with the MiniBooNE data~\cite{neutr_pi+}.} \label{fig2}
\end{figure}
\begin{figure}[t]
\includegraphics[width=\columnwidth]{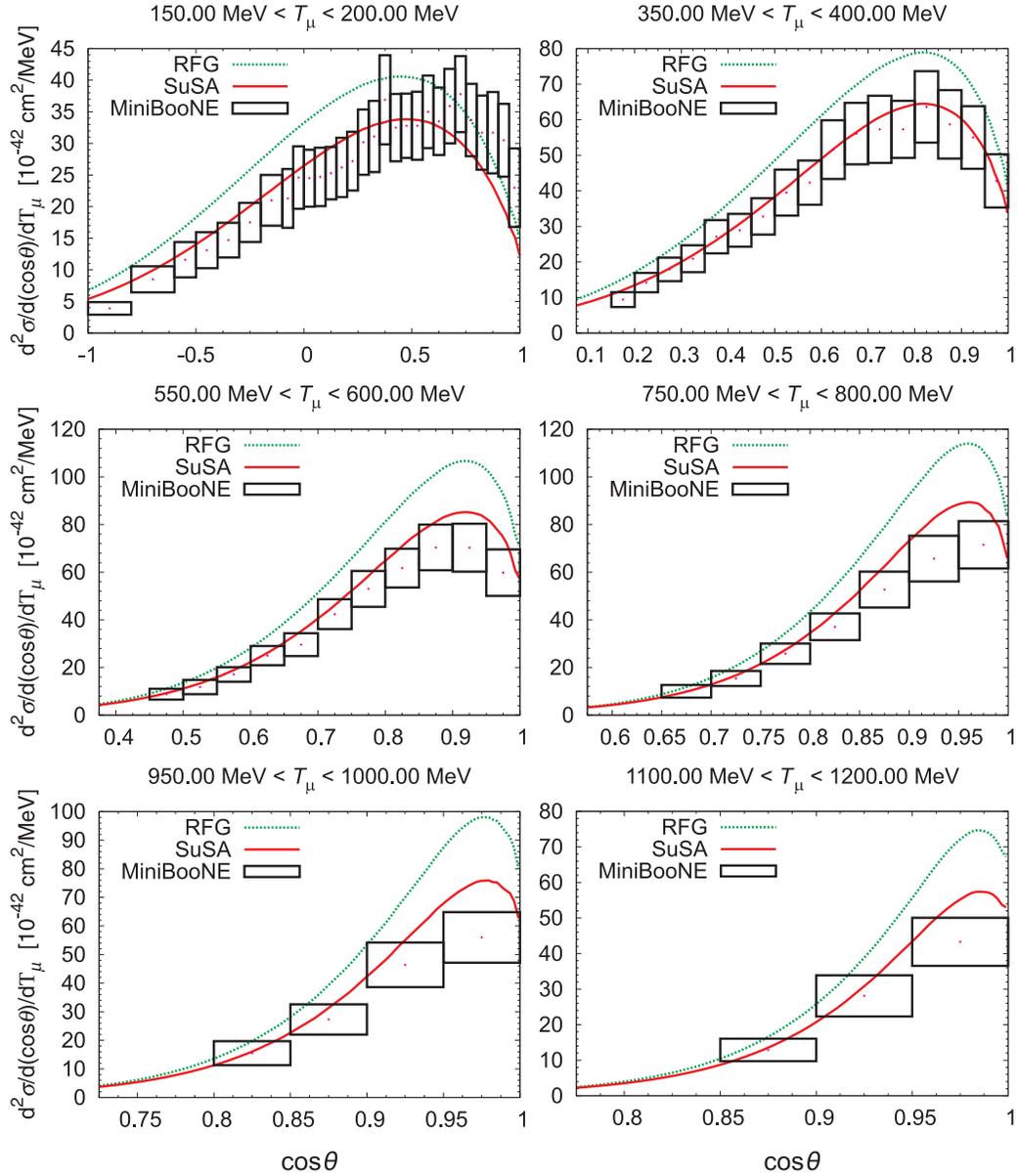}
\caption{(Color online) Same as Fig.~\ref{fig2} but the double-differential cross section averaged over the neutrino energy flux as a function of the scattering angle, are presented.} \label{fig3}
\end{figure}

\section{Results and discussion \label{sec:3}}

{\bc First we present RFG and SuSA predictions for the double-differential cross section for CC neutrino-induced $\pi^+$ production
on CH$_{2}$} averaged over the neutrino flux $\Phi(\epsilon_\nu)$, namely
\begin{equation}
\frac{d^2\sigma}{dT_\mu d\cos\theta}
= \frac{1}{\Phi_\text{tot}} \int
\left[ \frac{d^2\sigma}{dT_\mu d\cos\theta} \right]_{\epsilon_\nu}
\Phi(\epsilon_\nu) d\epsilon_\nu ,
\end{equation}
where $T_\mu$ and $\theta$ are correspondingly the kinetic energy and scattering angle of the outgoing muon, $\epsilon_\nu$ is the neutrino
energy and $\Phi_\text{tot}$ is the total integrated  $\nu_\mu$ flux factor for the MiniBooNE experiment ($\Phi_\text{tot} = 5.19\times 10^{-10}$
 [$\nu_\mu$/cm$^2$/POT]). The double-differential cross section averaged over the neutrino energy flux as a function of the muon kinetic energy
$T_\mu$ is presented in Fig.~\ref{fig2}. Each panel corresponds to a bin of $\cos\theta$. {\bc The PR2 parametrization has been considered. Results with the PR1 parameterization are about 5\% higher, that is a measure of the degree of uncertainty that we expect from the choice of the single-nucleon response for this reaction}. We compare the predictions of SuSA and RFG with the MiniBooNE data~\cite{neutr_pi+}. The
nuclear target has been considered as carbon and hydrogen in the mineral oil target. Fig.~\ref{fig3} shows SuSA and RFG predictions for
the double-differential cross section averaged over the neutrino energy flux as a function of the scattering angle at fixed
$T_\mu$ compared with data~\cite{neutr_pi+}.

Figs.~\ref{fig2} and~\ref{fig3} show a good agreement between data and the SuSA predictions for the flux-averaged double-differential cross sections. This applies to both parameterizations of the vector and axial form factors. RFG results have similar shape as SuSA ones but, as expected, they overestimate the data
to a large extent.

\begin{figure}[t]
  \centering
\includegraphics[width=0.8\columnwidth]{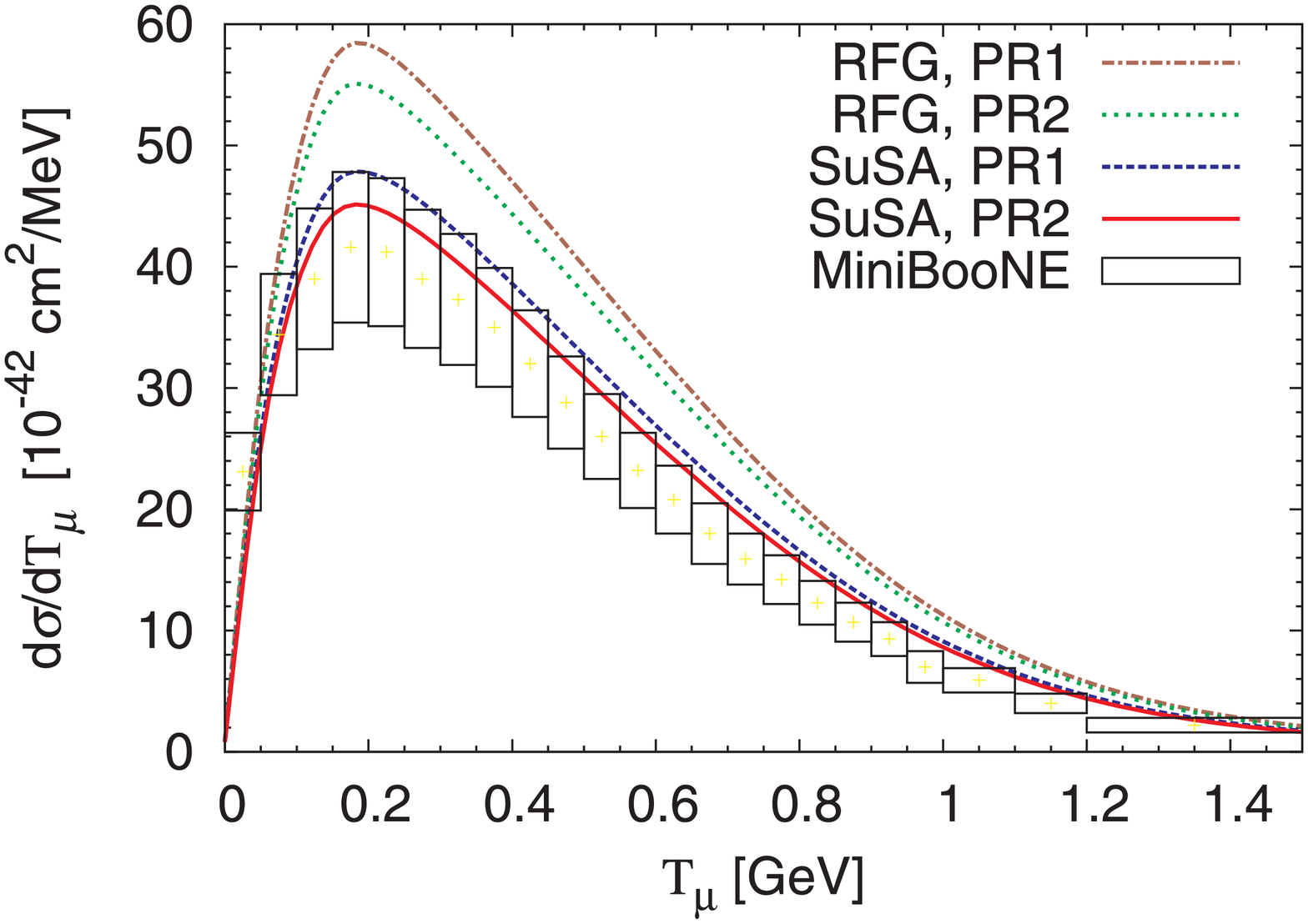}
\caption{(Color online) The ${d\sigma}/{dT_\mu}$ results obtained by integrating the flux-averaged double-differential
cross sections over $\cos\theta$ [Eqs.~(\ref{hiphop1})] are compared with the MiniBooNE data~\cite{neutr_pi+}.
For vector and axial form-factors two parameterizations, ``PR1''~\cite{neutr_Alvarez-Ruso99} and ``PR2''~\cite{neutr_Paschos04}, are used.} \label{fig4}
\end{figure}

In Fig.~\ref{fig4} are shown the results obtained by integrating the flux-averaged double-differential cross sections
over angle:
\begin{equation}
\left\langle\dfrac{d\sigma}{dT_\mu}\right\rangle=\dfrac{1}{\Phi_\text{tot}}
\!\!\int\!\!\Phi(\epsilon_\nu)\!\!\int
\!\!\left( \frac{d^2\sigma}{dT_\mu d\cos\theta} \right)_{\epsilon_\nu}\!\!\!\!d(\cos\theta)d\epsilon_\nu.\label{hiphop1}
\end{equation}

\begin{figure}[t]
  \centering
\includegraphics[width=0.8\columnwidth]{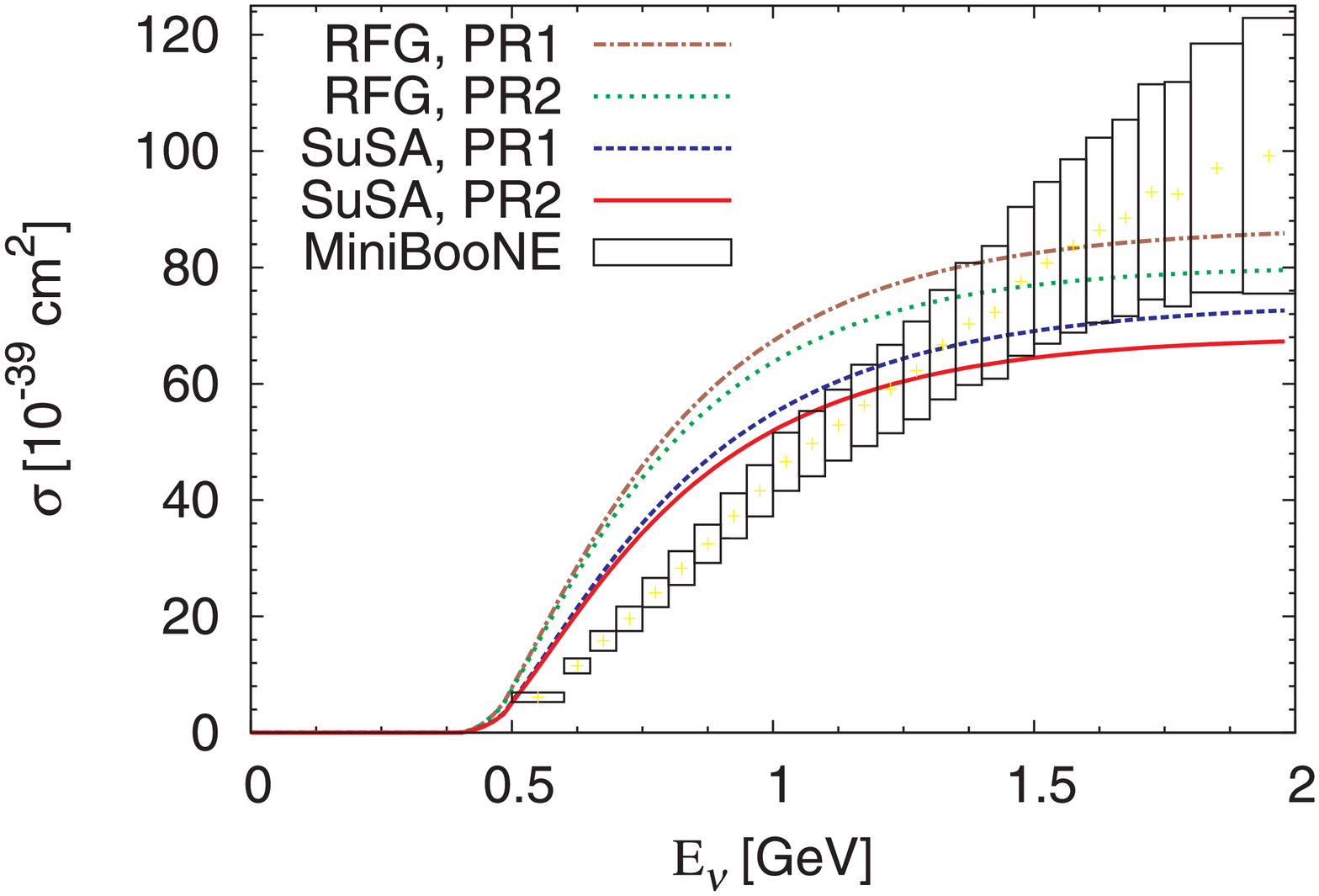}
\caption{(Color online) The total cross section for $\pi^+$ production are compared with the MiniBooNE data~\cite{neutr_pi+}.
For vector and axial form-factors two parameterizations, ``PR1''~\cite{neutr_Alvarez-Ruso99} and ``PR2''~\cite{neutr_Paschos04}, are used.} \label{fig5}
\end{figure}

The total cross section for $\pi^+$ production as a function of the neutrino energy along with the MiniBooNE data are displayed in Fig.~\ref{fig5}. Poorer agreement with data than for the flux-averaged cross sections presented in Figs.~\ref{fig2}--\ref{fig4} is clearly observed. The data seem to follow a more linear dependence with the energy up to $2$~GeV than the theory. However, before drawing definite conclusions, one has to consider on one side that the nuclear response extracted from the phenomenological electron data includes the whole (real) pion production strength (virtual pion contribution via MEC has already been removed) while in MiniBoone the data are sensitive only to the cases where a real pion is seen away from the nucleus. This means that if the real pion is produced and then absorbed during the final state, it will not be seen in the observed MiniBoone pionic cross-sections. On the other hand, the unfolding procedure used to extract the data of Fig.~\ref{fig5} {\bc is to some extent model dependent. Thus these data are less direct and we consider the comparison with the data of Figs.~\ref{fig2}--\ref{fig4} to be of more significance}.  {\bc It is worth mentioning some recent publications where the problems with the reconstruction of neutrino energy $E_{\nu}$ are discussed in more details~\cite{neutr_Martini12,neutr_Leitner10,neutr_Benhar09} and also the review by H. Gallagher, G. Garvey, and G.P. Zeller~\cite{neutr_Gallagher}, where the authors consider in depth neutrino-nucleus interactions (in the medium-energy regime, ${O}(1$~GeV$)$) in respect to the modern neutrino oscillation experiments.}

\begin{figure}[t]
  \centering
\includegraphics[width=0.8\columnwidth]{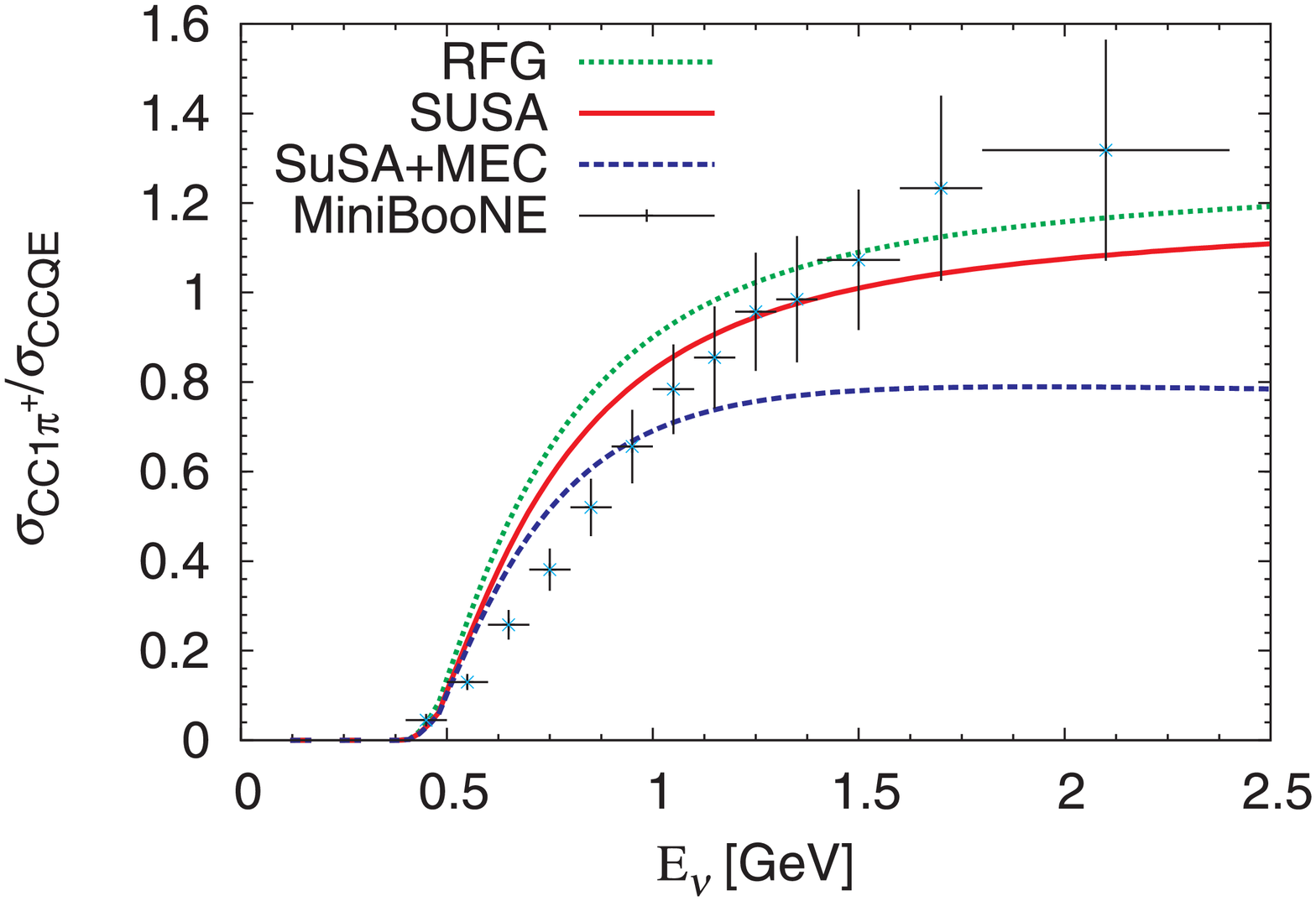}
\caption{(Color online) The results for CC1$\pi^+$ to CCQE cross section ratio are compared with MiniBooNE data
(corrected for final state interactions and rescaled for an isoscalar target)~\cite{neutr_ratio}.} \label{fig6}
\end{figure}

Figure~\ref{fig6} shows the ratio of CC1$\pi^+$ (CC single-pion production) to CCQE (CC quasielastic scattering) cross sections from SuSA, SuSA+MEC ($2p$--$2h$ meson-exchange current)~\cite{neutr_Amaro11}, and RFG approaches in comparison with the MiniBooNE data corrected for final state interactions. All these ratios have been rescaled to an isoscalar target~\cite{neutr_ratio}. The results are obtained on the basis of total cross sections for CC1$\pi^+$ (given in Fig.~\ref{fig5}) and CCQE~\cite{neutr_Amaro11}. A similar conclusion as the one in the previous figure could be drawn here. It seems that there is too much $\pi$ production strength below 1.2 GeV, and too little beyond that, compared to data.

{\bb Before concluding we would like to remind that the $\Delta$ scaling function used in our approach represents a good fit of electron scattering
data only in the region below the $\Delta$-peak, as shown in Fig.~\ref{fig1}. As anticipated, higher resonances and the tail of DIS come into play at high momentum and energy transfer. However, in the kinematical conditions of the MiniBooNE experiment the dominant contribution is associated to $\Delta$-excitation and therefore we only include the latter in our analysis. This is supported by the fact that inclusive electron scattering data in the same kinematical domain are reasonably well reproduced by using the pure $\Delta$-superscaling function (see Ref.~\cite{neutr_Amaro04})}.


\section{Conclusion \label{sec:4}}

Summarizing, in this paper we present results for the cross sections of neutrino-induced CC $\pi^+$ production obtained with the SuSA and RFG (shown as reference) models. The SuSA approach provides neutrino-nucleus cross-section predictions, based on the observed nuclear response to electron projectile and the universal character of the scaling function. Notice that SuSA predictions incorporate effects of final state interaction (FSI), the properties of the $\Delta$ resonance in the nuclear medium, {\bc both the contribution of coherent and incoherent production}, \emph{etc.}. The role of the FSI on the one-pion production has been considered for instance within the GIBUU transport model~\cite{neutr_Lalakulich11}, where it was shown that in order to reproduce the data, the total cross section obtained with FSI included has to be multiplied by a factor of $1.5$. Here we show that SuSA predictions are in good agreement with the MiniBooNE experimental data for pionic cross-section in the case of the flux averaged data, while some disagreement remains in the comparison to  unfolded neutrino energy data. Notice that the accordance between SuSA and data here is better than the one for the non-pionic case, where the model was found to underpredict the data unless meson exchange currents were explicitly included~\cite{neutr_Amaro11}. We conclude that the SuSA scaling function for the $\Delta$-region (extracted from electron scattering experiments) and its extension to neutrino processes {\bc may be very useful to predict} cross sections for neutrino-induced CC $\pi^+$ production, not relying on specific models.

\section*{Acknowledgements}
This work was partially supported by DGI (Spain) and FEDER funds under contracts FIS2011-28738-C02-01, FIS2008-01301, and FIS2011-23565, as well as the Bulgarian National Science Fund under contracts no. DO-02-285 and DID–02/16–17.12.2009, by Universidad Complutense de Madrid (UCM, 910059) and by the INFN-MICINN
collaboration agreement. M.V.I. is grateful for the warm hospitality given by the UCM and for financial support during his stay there from the
\emph{Centro Nacional de F\'isica de Part\'iculas, Astropart\'iculas y Nuclear} (CPAN) of Spain (Ref.: CSD2007-00042) and FPA2010-17142.
The authors would like to thank Rex Tayloe from the MiniBooNE collaboration for helpful discussions.


\begin{thebibliography}{100}

\bibitem{neutr_pi+} A.A. Aguilar-Arevalo \emph{et al.} (MiniBooNE Collaboration), Phys. Rev. D {83} (2011) 052007; A.A. Aguilar-Arevalo \emph{et al.} (MiniBooNE Collaboration), Phys. Rev. D {83} (2011) 052009.





\bibitem{neutr_sick80} I. Sick, D. B. Day, and J. S. McCarthy, Phys. Rev. Lett. {45} (1980) 871.

\bibitem{neutr_Ciofi} C. Ciofi degli Atti, E. Pace and G. Salm\`{e}, Phys. Rev. C {36} (1987) 1208; C. Ciofi degli Atti, E. Pace and G. Salm\`{e}, Phys. Rev. C {39} (1989) 259; C. Ciofi degli Atti, E. Pace and G. Salm\`{e}, Phys. Rev. C {43} (1991) 1155; C. Ciofi degli Atti, D.B. Day and S. Liuti, Phys. Rev. C {46} (1992) 1045; C. Ciofi degli Atti and S. Simula, Phys. Rev. C {53} (1996) 1689; C. Ciofi degli Atti and G. B. West, Phys. Lett. B {458} (1999) 447.

\bibitem{neutr_Day90} D. B. Day, J. S. McCarthy, T. W. Donnelly, and I. Sick, Annu. Rev. Nucl. Part. Sci. {40} (1990) 357.

\bibitem{neutr_Alberico90} W. M. Alberico, A. Molinari, T. W. Donnelly, E. L. Kronenberg, and J. W. Van Orden, Phys. Rev. C {38} (1988) 1801.

\bibitem{neutr_Barbaro98} M. B. Barbaro, R. Cenni, A. De Pace, T. W. Donnelly, and A. Molinari, Nucl. Phys. A {643} (1998) 137.

\bibitem{neutr_Donnelly99} T. W. Donnelly and I. Sick, Phys. Rev. Lett. {82} (1999) 3212; Phys. Rev. C {60} (1999) 065502.

\bibitem{neutr_Maieron02} C. Maieron, T. W. Donnelly, and I. Sick, Phys. Rev. C {65} (2002) 025502.

\bibitem{neutr_Barbaro04} M.~B.~Barbaro, J.~A.~Caballero, T.~W.~Donnelly, and C.~Maieron, Phys. Rev. C {69} (2004) 035502.

\bibitem{neutr_Antonov} A. N. Antonov, M. K. Gaidarov, D. N. Kadrev, M. V. Ivanov, E. Moya de Guerra, and J. M. Udias, Phys. Rev. C {69} (2004) 044321;
    A. N. Antonov, M. K. Gaidarov, M. V. Ivanov, D. N. Kadrev, E. Moya de Guerra, P. Sarriguren, and J. M. Udias, Phys. Rev. C {71} (2005) 014317; A. N. Antonov, M. V. Ivanov, M. K. Gaidarov, E. Moya de Guerra, P. Sarriguren, and J. M. Udias, Phys. Rev. C {73} (2006) 047302;
    A. N. Antonov, M. V. Ivanov, M. K. Gaidarov, E. Moya de Guerra, J. A. Caballero, M. B. Barbaro, J. M. Udias, and P. Sarriguren, Phys. Rev. C {74} (2006) 054603;
    A. N. Antonov, M. V. Ivanov, M. K. Gaidarov, and E. Moya de Guerra, Phys. Rev. C {75} (2007) 034319;
    M. V. Ivanov, M. B. Barbaro, J. A. Caballero, A. N. Antonov, E. Moya de Guerra, and M. K. Gaidarov, Phys. Rev. C {77} (2008) 034612.

\bibitem{neutr_Amaro04} J.~E.~Amaro, M.~B.~Barbaro, J.~A.~Caballero, T.~W.~Donnelly, A.~Molinari and I.~Sick, Phys. Rev.  C {71} (2005) 015501.

\bibitem{neutr_Jourdan96} J.~Jourdan, Nucl. Phys.  A {603} (1996) 117.

\bibitem{neutr_Amaro06} J.~E.~Amaro, M.~B.~Barbaro, J.~A.~Caballero and T.~W.~Donnelly, Phys. Rev.  C {73} (2006) 035503.


\bibitem{NQE2} C.~Maieron, J.~E.~Amaro, M.~B.~Barbaro, J.~A.~Caballero, T.~W.~Donnelly, C.~F.~Williamson, Phys. Rev. C {80} (2009) 035504.

\bibitem{neutr_Amaro10} J.~E.~Amaro, M.~B.~Barbaro, J.~A.~Caballero, T.~W.~Donnelly, C.~F.~Williamson,
Phys. Lett. B {696} (2011) 151.

\bibitem{neutr_Amaro11} J. E. Amaro, M. B. Barbaro, J. A. Caballero, T. W. Donnelly, and J. M. Udias, Phys. Rev. D {84} (2011) 033004.


\bibitem{Amaro:2011aa} J.~E.~Amaro, M.~B.~Barbaro, J.~A.~Caballero and T.~W.~Donnelly, arXiv:1112.2123 [nucl-th].

\bibitem{neutr_Umino95} Y. Umino and J. M. Udias, Phys. Rev. C {52} (1995) 3399.

\bibitem{neutr_Alvarez-Ruso99} L.~Alvarez-Ruso, S.~K.~Singh and M.~J.~Vicente Vacas, Phys. Rev. C {59} (1999) 3386.

\bibitem{neutr_Paschos04} E.~A. Paschos, J.-Y. Yu and M.~Sakuda, Phys. Rev. D {69} (2004) 014013.


\bibitem{neutr_Martini12} M. Martini, M. Ericson, G. Chanfray, {arXiv:1202.4745 [hep-ph]}.

\bibitem{neutr_Leitner10} T. Leitner and U. Mosel, Phys. Rev. C 81 (2010) 064614.

\bibitem{neutr_Benhar09} O. Benhar and D. Meloni, Phys. Rev. D 80 (2009) 073003.

\bibitem{neutr_Gallagher} H. Gallagher, G. Garvey, and G.P. Zeller, Ann. Rev. Nucl. Part. Sci 61 (2011) 355--378.

\bibitem{neutr_ratio} A.A. Aguilar-Arevalo \emph{et al.} (MiniBooNE Collaboration), Phys. Rev. Lett. {103} (2009) 081801.

\bibitem{neutr_Lalakulich11} O. Lalakulich, K. Gallmeister, T. Leitner, and U. Mosel, {arXiv:1107.5947v1 [nucl-th]}.

\end{thebibliography}
\end{document}